\documentclass[namedreferences,hyperref,optionalrh]{spr-sola}
\usepackage{graphicx}        
\usepackage{color}           

\chardef\us=`\_

\begin{document}

\begin{frontmatter}
\title{Observation of a large-scale filament eruption initiated by two small-scale erupting filaments pushing out from below}

\author[addressref={aff1,aff2},corref,email={ylsong@nao.cas.cn}]{\inits{Y.L.}\fnm{Yongliang}~\snm{Song}\orcid{0000-0002-9961-4357}}
\author[addressref={aff1,aff2,aff3},email={sjt@nao.cas.cn}]{\inits{J.T.}\fnm{Jiangtao}~\snm{Su}\orcid{0000-0002-5152-7318}}
\author[addressref={aff4}]{\inits{Q.M.}\fnm{Qingmin}~\snm{Zhang}\orcid{0000-0003-4078-2265}}
\author[addressref={aff1,aff2,aff3}]{\inits{M.}\fnm{Mei}~\snm{Zhang}\orcid{0000-0002-3141-747X}}
\author[addressref={aff1,aff2,aff3}]{\inits{Y.Y.}\fnm{Yuanyong}~\snm{Deng}\orcid{0000-0003-1988-4574}}
\author[addressref={aff1,aff2,aff3}]{\inits{X.Y.}\fnm{Xianyong}~\snm{Bai}\orcid{0000-0003-2686-9153}}
\author[addressref={aff1,aff2}]{\inits{S.}\fnm{Suo}~\snm{Liu}\orcid{0000-0002-1396-7603}}
\author[addressref={aff1,aff2}]{\inits{X.}\fnm{Xiao}~\snm{Yang}\orcid{0000-0003-1675-1995}}
\author[addressref={aff1,aff2}]{\inits{J.}\fnm{Jie}~\snm{Chen}\orcid{0000-0001-7472-5539}}
\author[addressref={aff1,aff2}]{\inits{H.Q.}\fnm{Haiqing}~\snm{Xu}\orcid{0000-0003-4244-1077}}
\author[addressref={aff6}]{\inits{K.F.}\fnm{Kaifan}~\snm{Ji}}
\author[addressref={aff1,aff2,aff3}]{\inits{Z.Y.}\fnm{Ziyao}~\snm{Hu}}
\address[id=aff1]{National Astronomical Observatories, Chinese Academy of Sciences, Beijing, 100101, China}
\address[id=aff2]{Key Laboratory of Solar Activity and Space Weather, National Space Science Center, Chinese Academy of Sciences, Beijing 100190, China}
\address[id=aff3]{School of Astronomy and Space Sciences, University of Chinese Academy of Sciences, Beijing 100049, China}
\address[id=aff4]{Key Laboratory of Dark Matter and Space Astronomy, Purple Mountain Observatory, Nanjing 210023, China}
\address[id=aff6]{Yunnan Observatories, Chinese Academy of Sciences, Kunming Yunnan 650216, China}

\runningauthor{Song et al.}
\runningtitle{Large-Scale Filament Eruption Initiated by Underlying  Ones}

\begin{abstract}
Filament eruptions often result in flares and coronal mass ejections (CMEs).  Most studies attribute the filament eruptions to their instabilities or magnetic reconnection.  In this study, we report a unique observation of a filament eruption whose initiation process has not been reported before. This large-scale filament, with a length of about 360 Mm crossing an active region, is forced to erupted by two small-scale erupting filaments pushing out from below. This process of multi-filament eruption results in an M6.4 flare in the active region NOAA 13229 on 25th February 2023.  The whole process can be divided into three stages: the eruptions of two active-region filaments F1 and F2; the interactions between the erupting F1, F2, and the large-scale filament F3; and the eruption of F3.  Though this multi-filament eruption occurs near the northwest limb of the solar disk, it produces a strong halo CME that causes a significant geomagnetic disturbance. Our observations present a new filament eruption mechanism, in which the initial kinetic energy of the eruption is obtained from and transported to by other erupting structures. This event provides us a unique insight into the dynamics of multi-filament eruptions and their corresponding effects on the interplanetary space.
    
\end{abstract}
\end{frontmatter}


\section{Introduction}
     \label{S-Introduction} 

Filaments are observed as dark elongated structures on the solar disk, in the H$\alpha$ and extreme ultraviolet (EUV) images. Over the limb, they are observed as bright structures and are called prominences. They are usually located along the magnetic polarity inversion lines (PILs), and are observed as the cooler and denser plasma, suspending in the dips of coronal magnetic structures. There are two typical models that describe the magnetic topology of filaments: the sheared arcade model and the flux rope model \citep[e.g.][]{Aulanier2002, Mackay2010, Guo2010, Xia2014, Jiang2014}. The typical plasma temperature and density of filaments are about 6000 K and $10^{11}$ cm$^{-3}$, respectively, which are about 100 times cooler and denser than their surrounding corona \citep[e.g.][]{Mackay2010, Priest2014}.

Filaments can be classified into three types, according to their locations, sizes and activities. The first type of filaments locate in the active regions, with each of their two legs rooted in or near the sunspots of different magnetic polarities.  This type of filaments are called {\it active-region filaments}. They are usually very active and easy to erupt. However, the scale of this type of filaments is usually not large, with their lengths confined within their locating active regions. The second type of filaments are called {\it quiescent filaments}. They often locate above the PILs of enhanced magnetic networks.  This type of filaments are relatively stable and not easy to erupt. But this type of filaments are usually much thicker and longer than {\it active-region filaments}. Their lengths can reach up to a scale of 1000 Mm \citep[e.g.][]{Priest2014, Chen2020}. The third type of filaments are called {\it intermediate filaments}. They are those between the former two types. They usually locate outside active region cores, with their two legs are rooted at the boundary of the active region or one leg at the active region boundary and the other in the quiet Sun \citep{Priest2014}.
 
Filaments play a significant role in solar activity. Most of their eruptions lead to solar flares and coronal mass ejections (CMEs). There are mainly two types of physical mechanisms to explain filament eruptions. One is the magnetic reconnection. This includes flux emergence, tether-cutting, breakout, and catastrophic models \citep[e.g.][]{Antiochos1999, Chen2000, Lin2000, Moore2001, Zhang2020}.  The other is ideal magnetohydrodynamic (MHD) instabilities. This includes kink instability, torus instability, and disturbances on filaments by other small-scale activities \citep[e.g.][]{Hood1981, Ji2003, Fan2005, Torok2005, Demoulin2010, Chen2018ApJ}.

For some active regions, the magnetic topology could be very complicated \citep[e.g.][]{Yan2015, Xu2020, Zheng2021, Li2022ApJ}. Several filaments can co-exist simultaneously or one filament can have several branches \citep[e.g.][]{Chandra2011, Liu2012, Cheng2014, Zhu2015, Hou2018, Su2018ApJ, Qiu2020}. One well-known model is the so-called ``double-decker" model \citep{Liu2012, Kliem2014}. It refers to the filament that consists of two vertically-distributed segments, of two magnetic flux ropes (MFRs) or of one MFR sitting over a sheared arcade magnetic structure. It is often used to explain the  partial eruption of the filament \citep[e.g.][]{Zhu2014, Zhang2015, Awasthi2019, Hou2023, Sun2023}.  Observations also show that two different filaments can make up a “double-decker” configuration \citep[e.g.][]{Zhu2015, Hu2022}. 

When an active region possesses two or more filaments simultaneously, the magnetic topology is usually very complicated. Collisions and interactions usually occur between these filaments, particularly when one of them begins to erupt \citep[e.g.][]{Linton2001, Su2007, Kumar2010, Torok2011, JiangYC2014, Zhu2015}. However, such cases are very rare in observations, especially for magnetic systems with more than two filaments.

In this study, we present a very special case where two erupting active-region filaments push the above large-scale filament (the third) out to eruption.  This event gives us a unique insight into the dynamics of the filaments during their eruptions. The paper is organized as follows. Section 2 describes the data. Section 3 presents the analysis and results.  And Section 4 gives the summary and discussion.

\section{Data}
\label{S-Obs}

The multi-filament eruption we studied occurred on 25th February 2023, in the northwest of the solar disk and was very close to the solar limb (see Fig. \ref{fig1}).  We carry out a comprehensive analysis using the data from both space- and ground-based telescopes. The {\it Atmospheric Imaging Assembly} (AIA: \citealp{Lemen2012}) on board the {\it Solar Dynamics Observatory} (SDO) can provide multi-wavelength images of the solar atmosphere. The time cadences for the extreme ultraviolet (EUV) passbands and the ultraviolet (UV) passbands are 12 s and 24 s, respectively.  The spatial sampling is about $0.6^{\prime\prime}$ per pixel. The {\it Extreme-ultraviolet Imager} (EUVI: \citealp{Wuelser2004}) on board the {\it Solar Terrestrial Relations Observatory Ahead} spacecraft (STEREO-A: \citealp{Kaiser2008}) also has good observational data for this event. The EUVI provides four-passband observations, including 171 \AA, 195 \AA,  284 \AA, and  304 \AA, with a spatial pixel size of about $1.6^{\prime\prime}$. We also use the observations of the Ne {\footnotesize VII} 46.5 nm spectral line, obtained by the {\it Solar Upper Transition Region Imager} (SUTRI: \citealp{Bai2023}) on board the {\it Space Advanced Technology demonstration satellite series} (SATech-01). SUTRI can image the upper transition region with a temperature of about 0.5 MK \citep{Tian2017}. The spatial pixel size and time resolution of SUTRI are about $1.2^{\prime\prime}$ and 30 s, respectively.

The {\it Advanced Space-based Solar Observatory} (ASO-S: \citealp{Gan2019}), launched on 9th October 2022, is the first Chinese space-borne comprehensive solar observatory. Its scientific objective is to study the relationship between the solar magnetic field and solar eruptions. Three instruments are deployed on ASO-S: the {\it Full-disk vector MagnetoGraph} (FMG: \citealp{Deng2019}), {\it Ly$\alpha$ Solar Telescope} (LST: \citealp{Chen2019, Li2019}) and {\it Hard X-ray Imager} (HXI: \citealp{Zhang2019}). In this study, we mainly use the magnetograms taken by FMG. It observes the Sun at the wavelength of Fe {\footnotesize I}  5324.19~\AA\ with a 4 K$\times$4 K CMOS camera. The spatial pixel size is about $0.5^{\prime\prime}$. Normally, it takes the line-of-sight (LOS) magnetograms with a cadence of 30 s, and for the vector magnetograms the cadence is 120 s.

The H$\alpha$ images taken by the {\it Chinese H$\alpha$ Solar Explorer} (CHASE: \citealp{Li2022}) and the {\it Global Oscillation Network Group} (GONG: \citealp{Harvey2011}) are also used. CHASE provides high-quality raster scanning observations of the full Sun with the spectral ranges of 6559.7 - 6565.9~\AA (H$\alpha$) and 6567.8 - 6570.6~\AA (Fe {\footnotesize I}). The spatial, temporal, and spectral resolutions are about $1^{\prime\prime}$, 1 minute, and 0.072~\AA, respectively. In order to show the CME, the data from the {\it Large Angle Spectroscopic Coronagraph} (LASCO: \citealp{Brueckner1995}) on board the {\it Solar and Heliospheric Observatory} (SOHO: \citealp{Domingo1995}) and COR2 coronagraph on board STEREO-A are also used.

\section{Analysis and Results} 
      \label{S-general}      

\begin{figure}    
\centerline{\includegraphics[trim=1.0cm 0.0cm 0.0cm 0.0cm, width=1.08\textwidth]{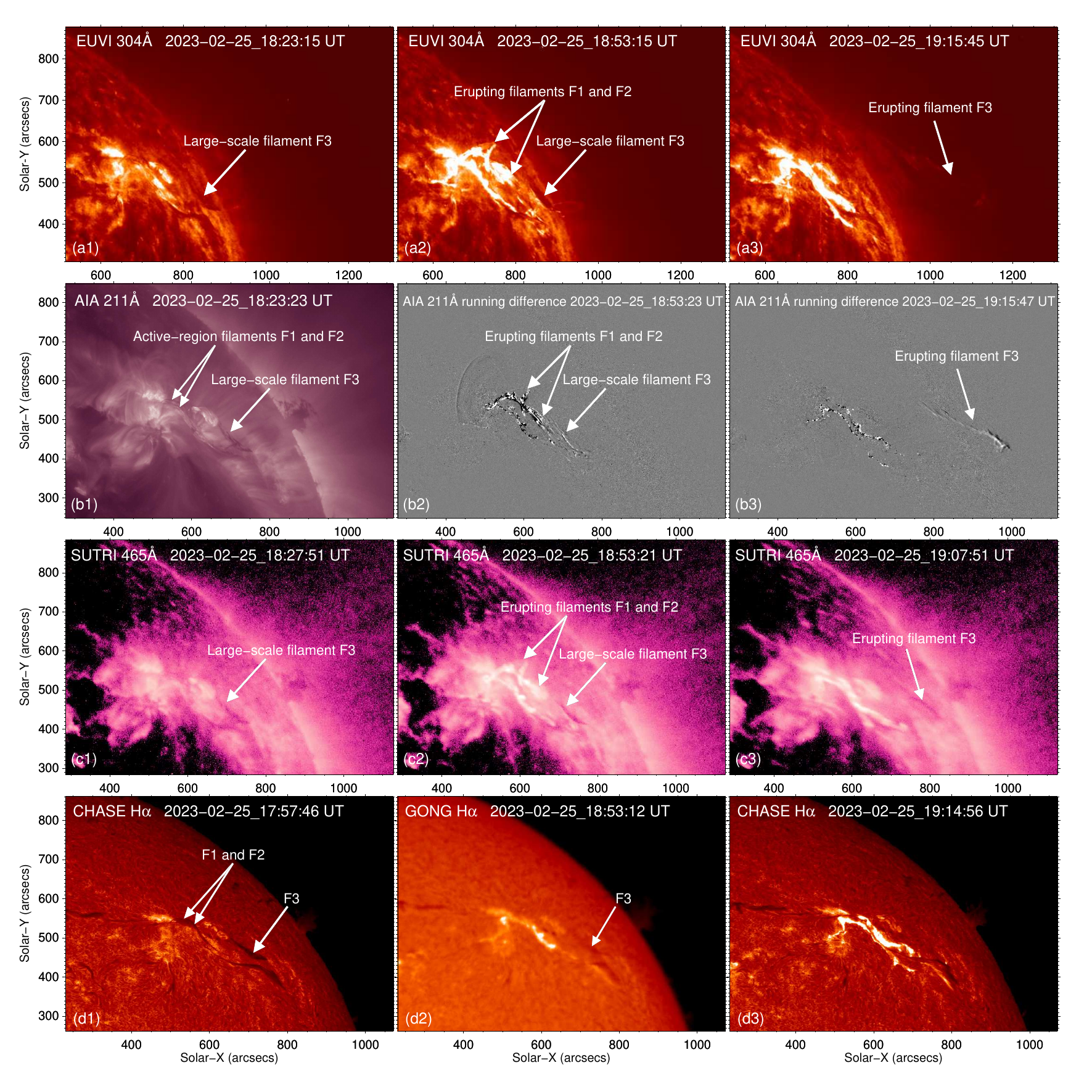}}
\small
 \caption{Overview of the filament eruptions in active region NOAA 13229. Panels a1 -- a3, b1 -- b3, c1 -- c3 and d1 -- d3 show the STEREO-A/EUVI 304~\AA , SDO/AIA 211~\AA  ~and the running difference, SUTRI 465~\AA , CHASE and GONG H$\alpha$ images, respectively. From left to right we show the pre-eruption, active-region filaments F1 and F2 eruptions, and large-scale filament F3 eruption being pushed out by the erupting F1 and F2. (animations for this multi-filament eruption observed by EUVI 304~\AA ~ and AIA 211~\AA~ running difference are available online.) }
\label{fig1}
\end{figure}

\begin{figure}    
\centerline{\includegraphics[trim=2.0cm 1.0cm 1.0cm 1.0cm, width=1.08\textwidth]{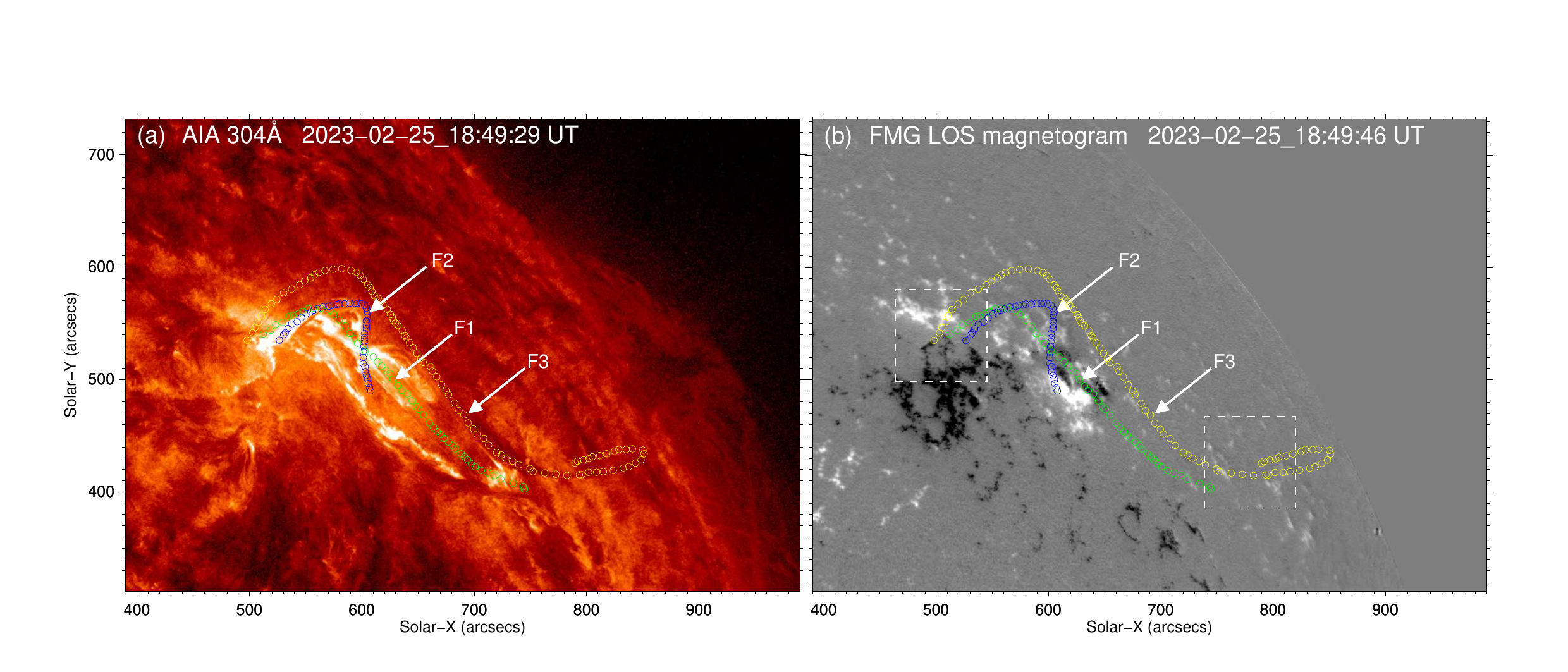}}
\small
        \caption{The configuration of the three filaments in NOAA 13229. Left panel a and right panel b show the locations of the filaments, overlaid on the AIA 304~\AA ~image and the ASO-S/FMG line-of-sight (LOS) magnetogram, resepctively, at 18:49 UT, when F1 and F2 are already erupting and F3 is still stable. Here all three filaments are outlined by circles but with different colors. Two white dashed boxes in panel b mark the foot areas of F3, which will be shown in Figure \ref{fig8}.}
\label{fig2}
\end{figure}

\begin{figure}    
\centerline{\includegraphics[trim=1.0cm 0.0cm 0.0cm 0.0cm, width=1.08\textwidth]{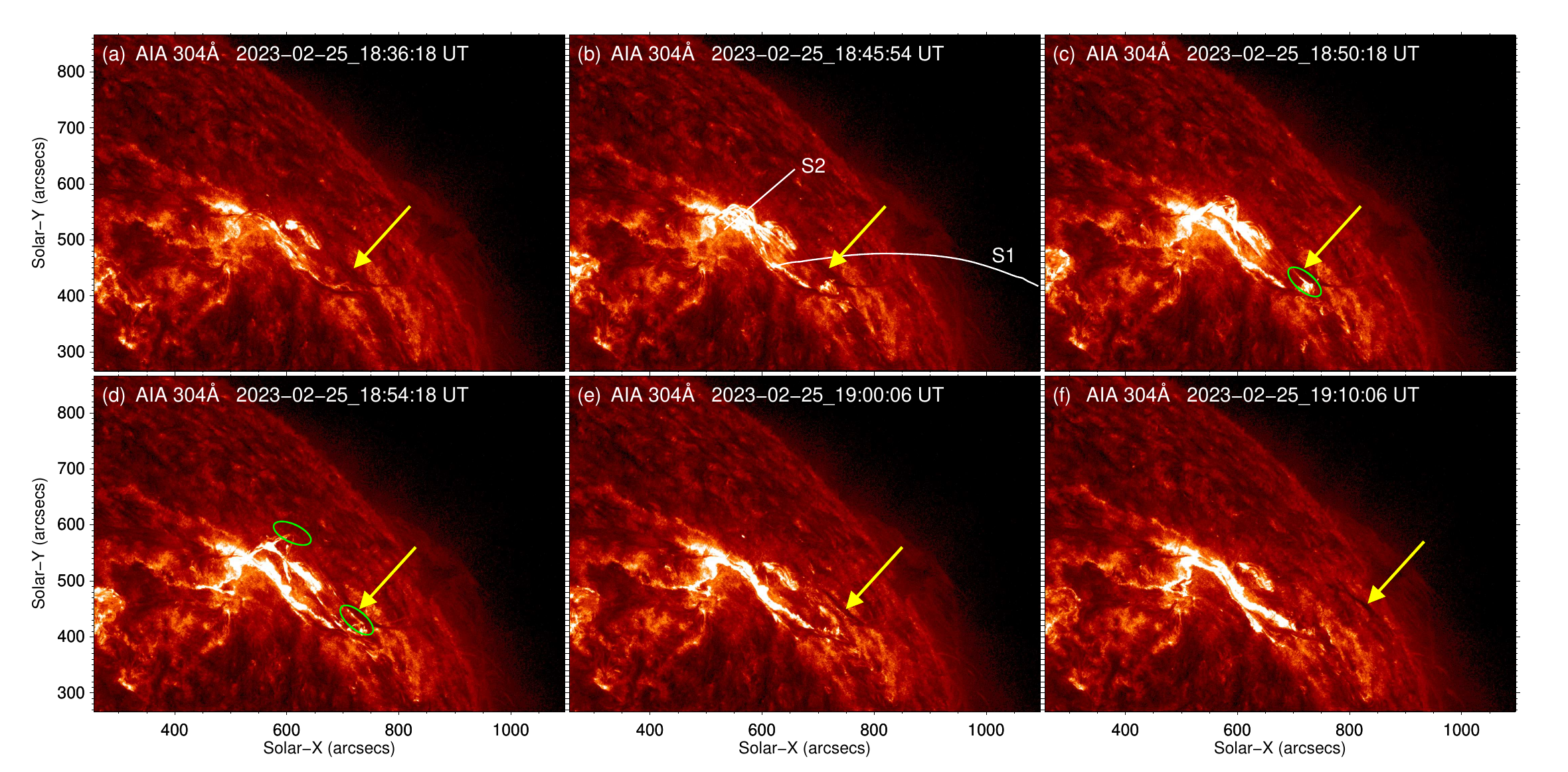}}
\small
        \caption{Interactions between active-region filaments F1, F2, and large-scale filament F3. Yellow arrows mark the F3 at different times.  Green elliptical circles indicate the positions where F1 and F2 collide with F3 and subsequently begin to push it.  Slits S1 and S2 correspond to the eruptions of F1 and F2 and their interactions with F3, respectively, time-distance maps of which will be shown in Figure \ref{fig5}. (an animation for this figure is available online.)   }
\label{fig3}
\end{figure}

\begin{figure}    
\centerline{\includegraphics[trim=1.0cm 0.0cm 0.0cm 0.0cm, width=1.08\textwidth]{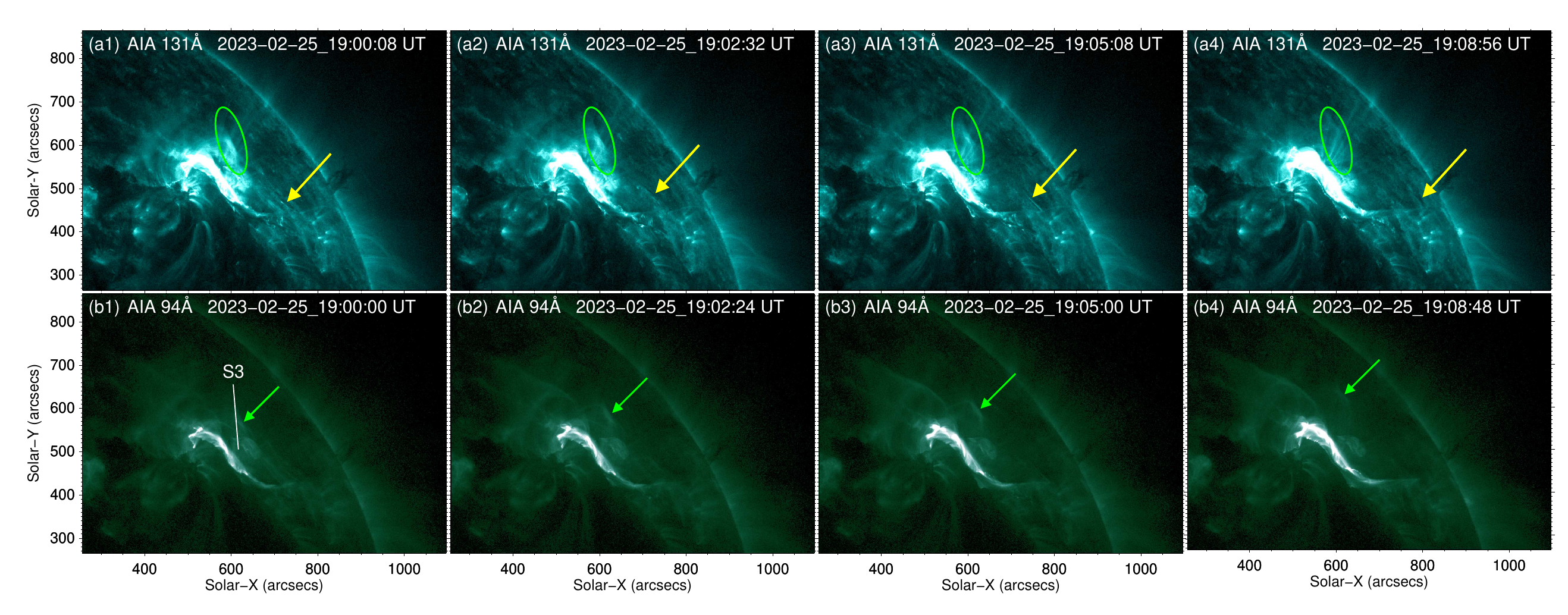}}
\small
        \caption{Interaction between filaments F2 and F3, from 19:00 UT to 19:10 UT.  Upper a1 --a4 and bottom b1 --b4 panels show the AIA high-temperature passbands 131 \AA ~and 94 \AA~ images, respectively.  In the upper panels, yellow arrows  mark the erupting F3.  Green elliptical circles indicate the positions where F2 pushes F3 with a bright structure moving upwards. Green arrows in the bottom panels point to the moving bright structure. Slit S3 is used to show the motion of the bright structure, time-distance maps of which will also be shown in Figure \ref{fig5}, together with S1 and S2.  (an animation for this figure is available online.)  }
\label{fig4}
\end{figure}

\begin{figure}    
\centerline{\includegraphics[trim=1.0cm 1.5cm 0.0cm 1.0cm, width=0.98\textwidth]{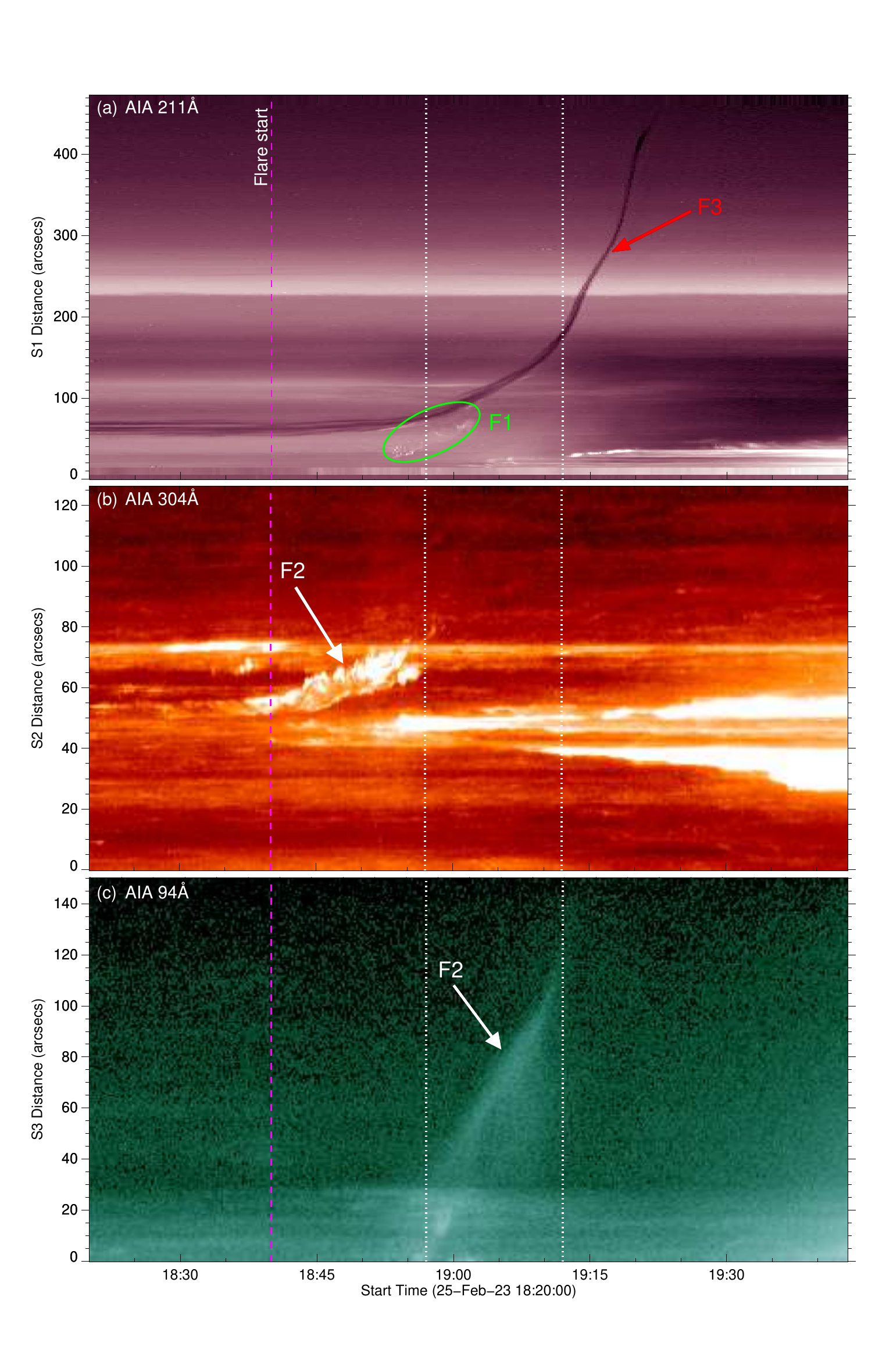}} 
\small
        \caption{Time-distance maps of the three slits.  From top to bottom, they are of S1, S2, and S3, using the AIA 211~\AA , 304~\AA ~and 94~\AA ~data, respectively. The positions of slits S1, S2, and S3 can be seen in Figures \ref{fig3} and \ref{fig4}.  The vertical magenta dashed lines outline the flare start time. The green elliptical circle in panel a indicates the eruption of F1, associated with expansion, rising and rotating motions, and then pushing F3 to eruption. White arrows in panel b point to the eruption of F2 and its interaction with F3 in panel c.  Red arrow marks the evolution of F3. The two vertical white dotted lines correspond to the period when F2 is pushing F3.  }
\label{fig5}
\end{figure}

\begin{figure}    
\centerline{\includegraphics[trim=1.0cm 0.0cm 0.0cm 0.0cm, width=0.98\textwidth]{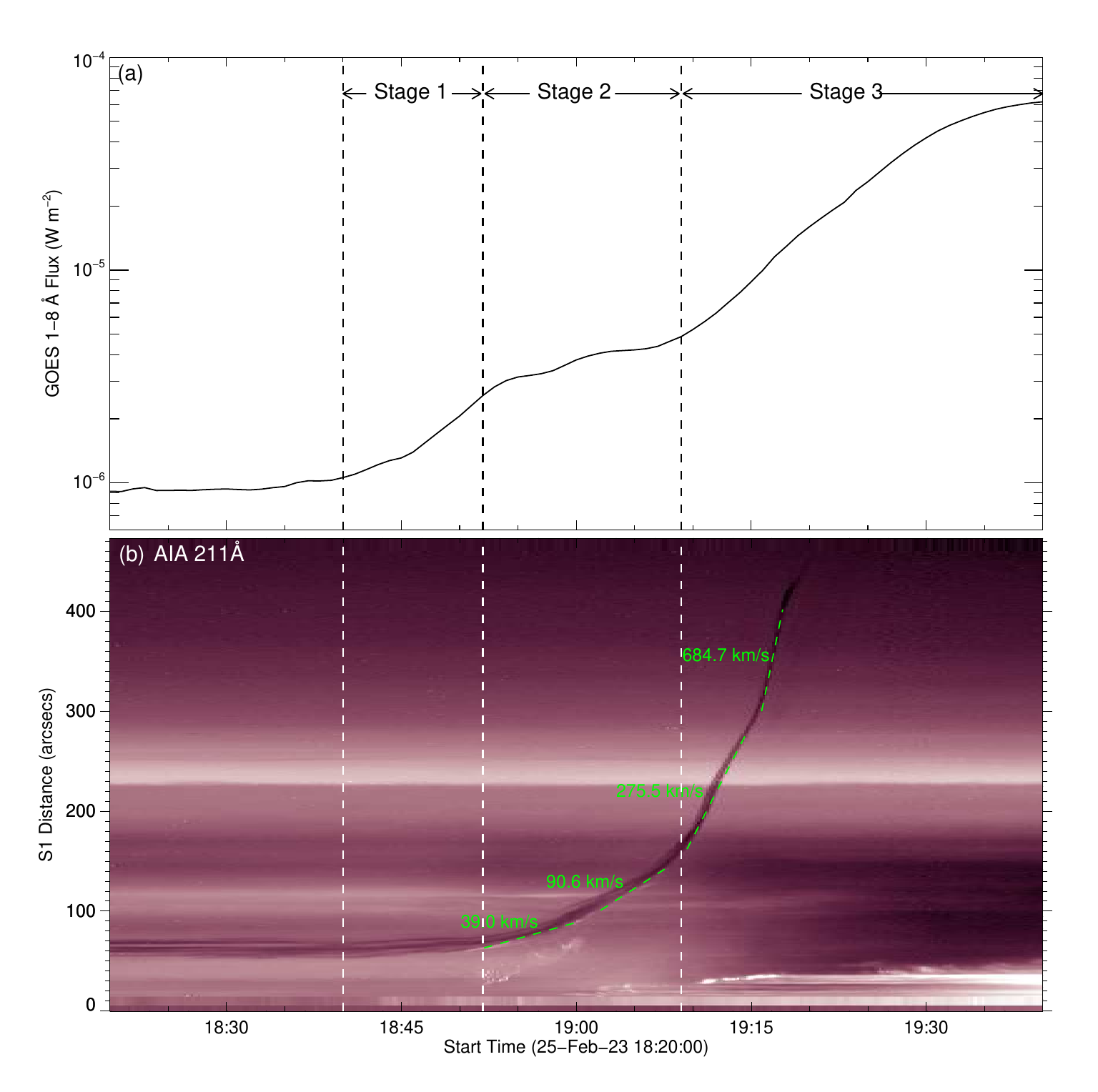}}
\small
        \caption{The three stages of the eruption. The upper panel shows GOES X-ray 1-8 \AA ~flux. The bottom is the time-distance plot of slit S1 using AIA 211~\AA ~data. Three stages can be seen that correspond to the eruptions of the two active-region filaments F1 and F2 (Stage 1), the interactions between both F1 and F2 with the large-scale filament F3 (Stage 2), and the eruption of F3 (Stage 3).  }
\label{fig6}
\end{figure}

\begin{figure}    
\centerline{\includegraphics[width=0.88\textwidth]{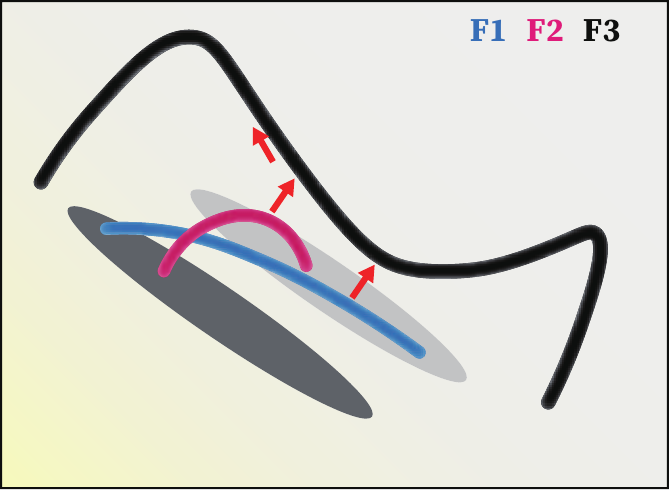}}
\small
        \caption{A cartoon that shows the interactions between the two active-region filaments F1 and F2 with the large-scale filament F3.  Red arrows indicate the directions in which F1 and F2, respectively, collide with F3 and subsequently push F3 to eruption. The grey and dark ellipses outline the locations of positive and negative magnetic fields in the active region.  }
\label{fig7}
\end{figure}

Figure \ref{fig1} is the overview of the multi-filament eruptions with the observations of the EUVI 304~\AA , AIA 211~\AA, SUTRI 465~\AA, CHASE and GONG H$\alpha$. Before the eruption, two sets of filaments can be seen clearly: the low-lying filaments F1 and F2, and the large-scale filament F3, with the last one more higher up in the solar atmosphere. These two sets of filaments can be easily distinguished in the H$\alpha$ observation (Fig. \ref{fig1}d1). However, it is a challenge to determine the filaments F1 and F2 only through H$\alpha$ images. Employing the AIA 211~\AA~ image we can more clearly see these two low-height filaments F1 and F2 (Fig. \ref{fig1}b1). Filaments F1 and F2 appear to intertwine with each other.

From the central panels of Figure \ref{fig1}, we can see that when F1 and F2 have already erupted and even when they have already risen to a somewhat higher height, F3 was still and remained stable. There was a significant height difference between the two sets of filaments before their eruptions. During the eruptions of F1 and F2, it is evident that their left legs are intertwined (Fig. \ref{fig1}b2). The flare ribbons are just located below F1 and F2. Finally, the large-scale filament F3 is pushed out to eruption by the erupting F1 and F2, and then we see two much longer flare ribbons (Fig. \ref{fig1}d3).

Figure \ref{fig2} shows the configuration of the three filaments, during the eruptions of F1 and F2 while F3 is still stable. We can see that F1 and F2 are two active-region filaments in NOAA 13229. F1 is located along the PIL of the active region, and F2 just crosses the PIL and rides above F1. Their left legs are very close and seem to be intertwined. F3 is a large-scale filament that crosses the active region with a length of about 360 Mm.  Its left leg is rooted in the network of negative magnetic polarity, at the northern edge of the active region.  Its right leg is rooted in a quiet Sun region, far away from the active region. Most part of this filament is nearly parallel to the F1, above the PIL of the active region.

Based on these observations, we regard filaments F1 and F2 as one magnetic system that coincides with the ``double-decker" model \citep{Liu2012}. We regard filament F3 as another independent magnetic system, due to its larger scale and higher altitude, relative to F1 and F2. We see in this event, the energy is first released by the lower magnetic system, that is, in the F1 and F2 eruptions, then by the upper magnetic system, that is,  in the F3 eruption.

Figure \ref{fig3} shows the interactions between the active-region filaments F1, F2, and the large-scale filament F3 from 18:36 UT to 19:10 UT. The associated flare starts at 18:40 UT. We see that the two filaments F1 and F2 are activated and begin to rise.  At the early stage, both filaments are seen as very bright all along (see Fig. \ref{fig3}b). It should come from the interchange magnetic reconnection between two filaments due to their special configuration. When F1 rises, it is accompanied by a slight expansion and rotating motion in the early stage about from 18:38 UT to 18:45 UT.  At about 18:50 UT, F1 first collides with F3 and begins to push it rising. We see a bright emission below F3, as marked by the green elliptical circle in Fig. \ref{fig3}c. At the same time, F2 is still rising and F3 begins to rise slowly. About four minutes later, at 18:54 UT, F2 collides with the left ``shoulder'' of F3 and begins to push F3 (ee the left green elliptical circle in Fig. \ref{fig3}d). Then, F3 rises much faster and the flare ribbons elongate.

The interaction between F2 and F3 is more noticeable in AIA 131 \AA ~and 94 \AA ~observations (Figure \ref{fig4}). It may be due to the high-temperature but low-density plasma in the left shoulder of F3 where F2 collides with it. We can see a bright structure sliding upwards smoothly (see the green elliptical circles in Fig. \ref{fig4}{a1--a4 and green arrows in Fig. \ref{fig4}b1--b4). This sliding motion persists for more than 10 minutes, indicating that F2 is pushing F3 to erupt. Simultaneously, we see that the large-scale filament F3 erupts more rapidly with the velocity increasing from about 39.0 km/s to about 90.6 km/s (see Fig. \ref{fig6}b).

Figure \ref{fig5} is the time-distance maps for slits S1, S2, and S3, where the filaments eruptions can be seen clearly. The flare starts at 18:40 UT. At the early stage of the flare, F1 and F2 erupt and rapidly rise but F3 remains stable (Fig. \ref{fig5}a and b). At around 18:50 UT, F1 first encounters F3 and begins to push it upwards. Then, we see that F3 begins to rise slowly. There is a slight brightening at the lower boundary of F3. In this process, we do not see very evident or strong reconnection between F1 and F3. This may be due to the fact that the two filaments run nearly parallel to each other along the PIL of the active region and with the same magnetic direction (Fig. \ref{fig2}). As pointed out by \citet{Linton2001}, magnetic reconnection between two flux tubes significantly depends on their relative orientation. However, small-scale and weak reconnections could still occur between the magnetic field lines in the outer shells of two filaments, since the magnetic field could still make an angle because of a helical or sheared structure. This slight brightening may come from the heating of the plasma at the border of filaments, resulting from the collision, squeezing and weak magnetic reconnection between the two filaments. The interaction between F1 and F3 lasts about 15 minutes.

F2 collides with F3 at around 18:55 UT (Fig. \ref{fig5}b and c). The interaction between F2 and F3 also lasts about 15 minutes. When only F1 pushes, F3 rises very slowly.  However, when F1 and F2 simultaneously push F3 from about 18:56 UT to 19:03 UT, we see that F3 begins to rise rapidly. Finally, after the interactions with F1 and F2, the large-scale filament F3 is ejected at a higher speed, possibly with a conversion of its own non-potential magnetic energy into kinetic energy together with that gained from the interactions with F1 and F2 in the form of pushing.

Figure \ref{fig6} shows the corresponding relationship between the GOES 1-8~\AA ~X-ray flux and the eruption of filament F3. Obviously, the development of the event can be divided into three stages. In the first stage from 18:40 UT to 18:52 UT, we see that the GOES X-ray flux increases rapidly. At this stage, the filaments F1 and F2 erupt but F3 remains stable.

The second stage is from 18:52 UT to 19:09 UT. During this period, we see a gradual deceleration in the increase of GOES X-ray flux. The eruptions of F1 and F2 are interrupted by the large-scale filament F3.  The interactions with the above large-scale filament F3 decrease the ejection velocities of the erupting F1 and F2.  Part of the kinetic energy of F1 and F2 is transferred to F3. At this stage, it is evident that F3 undergoes two distinct acceleration phases. In the first phase, that is, during its interaction with F1, its rising velocity increases from 0 to about 39.0 km/s. In the second phase, when it is interacting with both F1 and F2, its rising velocity increases from about 39.0 km/s to about 90.6 km/s.

The third stage is from 19:09 UT to the end of the flare. After interaction with F1 and F2, the large-scale filament F3 escapes the Sun at a significantly higher speed, reaching about 684.7 km/s. The GOES X-ray flux increases rapidly again and finally reaches the level of M6.4.

\begin{figure}    
\centerline{\includegraphics[trim=1.0cm 1.0cm 1.0cm 1.0cm, width=0.98\textwidth]{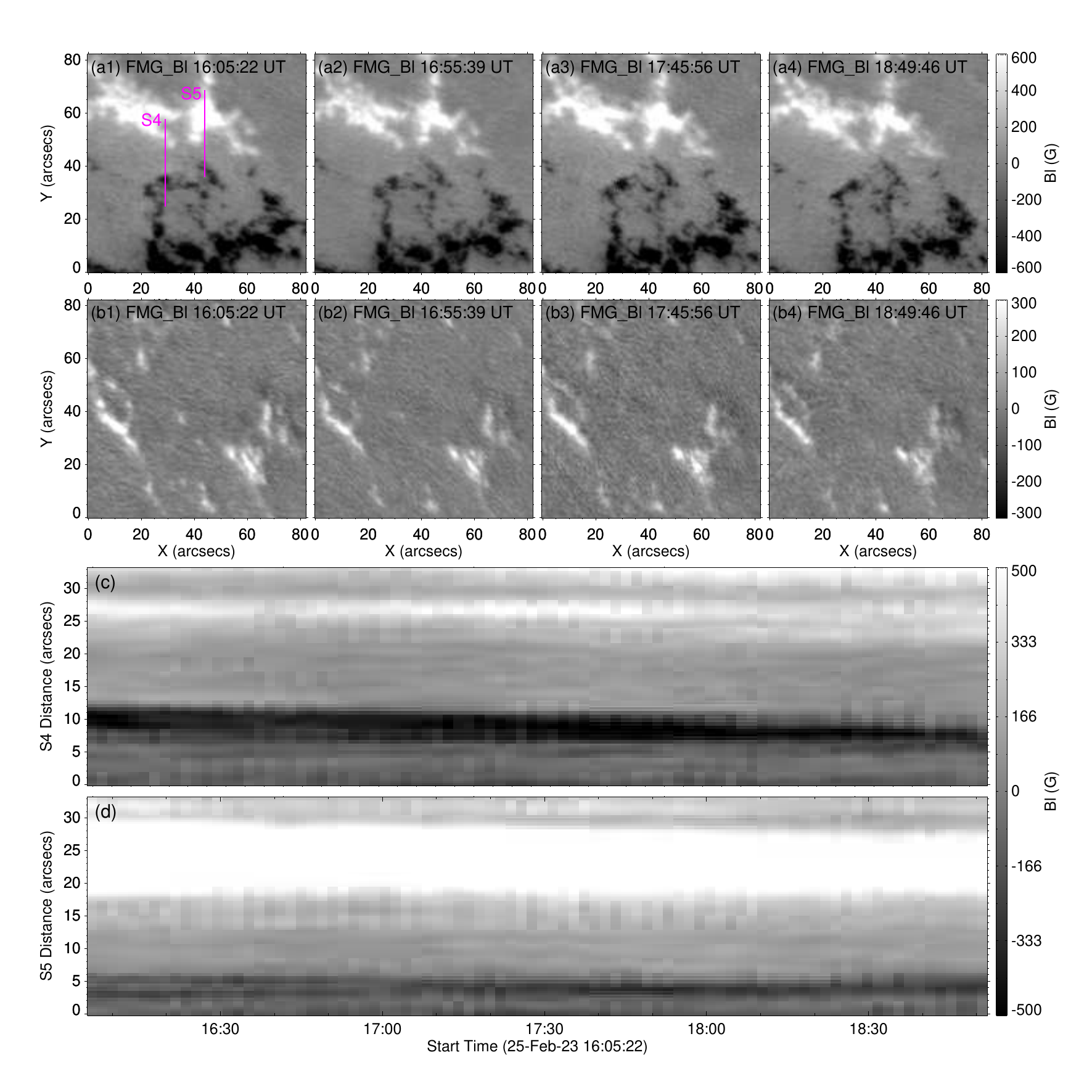}}
\small
        \caption{Temporal evolutions of the LOS magnetic field in the two regions where filament F3 is rooted as observed by ASO-S/FMG. The two regions are marked by white dashed boxes in Figure \ref{fig2}b. Panels a1 -- a4 and b1 --b4 show the left and right foot areas, respectively. Slits S4 and S5 are used to show the motions of different magnetic field polarities, time-distance maps of which are shown in panels c and d, respectively.}
\label{fig8}
\end{figure}

\begin{figure}    
\centerline{\includegraphics[width=0.98\textwidth]{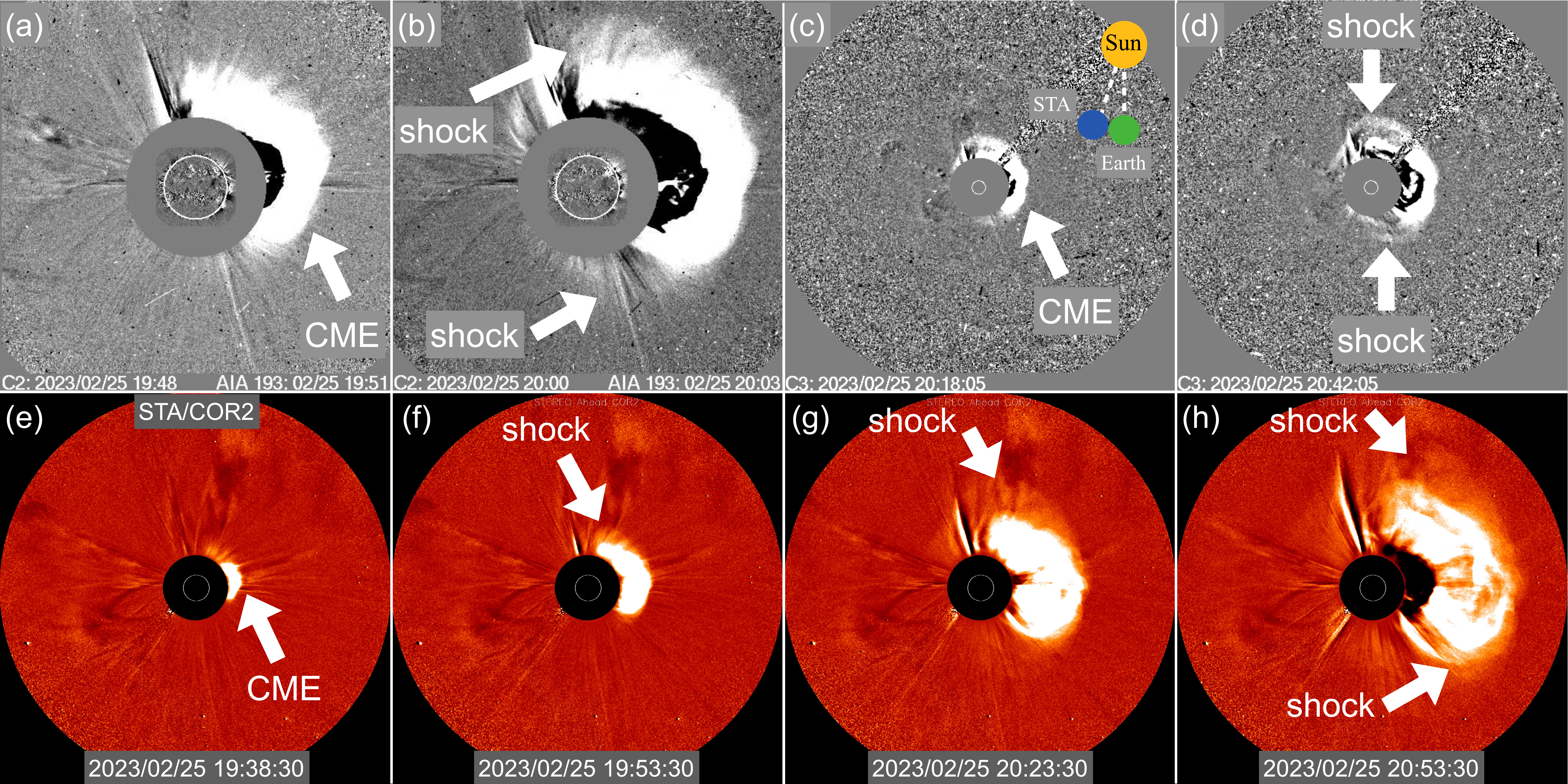}}
\small
        \caption{The CME resulting from the filament eruptions.   (a)-(b) and (c)-(d): LASCO/C2 and LASCO/C3 running difference images.  (e)-(h): STEREO-A/COR2 images.   The positions of STEREO-A (STA) satellite, Earth, and Sun are shown in the top right corner of panel c.  The separation angle of STA with respect to the Sun-Earth line is $12.85^\circ$. The CME and CME-driven shock are pointed by white arrows.  }
\label{fig9}
\end{figure}

\begin{figure}    
\centerline{\includegraphics[width=0.98\textwidth]{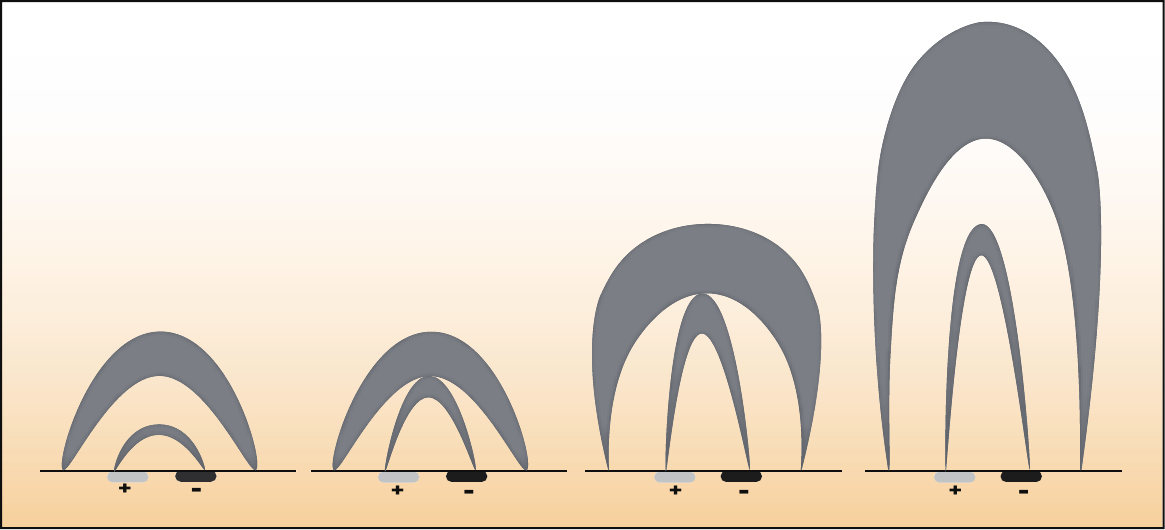}}
\small
        \caption{A schematic diagram that shows the large-scale filament eruption initiated by small-scale active-region filaments pushing out from below.  “+" and “-" denote the signs of magnetic polarities in the active region.  }
\label{fig10}
\end{figure}

Figure \ref{fig7} is a cartoon that shows the interactions between the filaments F1 and F2 with F3.  F1 and F2 are two active-region filaments located above the PIL with a ``double-decker" configuration \citep{Liu2012}.  F1 is longer and makes a relatively smaller angle with the PIL. F2 rides over F1 and has a relatively larger angle with the PIL. F3 is a large-scale filament crossing over the active region   with its axis nearly parallel to the PIL. F1 and F2 erupt almost simultaneously.  During the eruption, interchange reconnections occur between F1 and the two legs of F2. Both filaments rapidly rise and F1 is also accompanied by an expansion and rotational motion.
F1 is the first one that collides with F3 and begins to push F3, since the lower part of F3 is situated just above F1. It is about five minutes later that F2 collides with the left shoulder of F3 and begins to push F3 upwards along the shoulder. At last F3 is pushed out to eruption causing a large CME. F1 probably erupts successfully. But F2 is a failed eruption (see Fig. \ref{fig1}, \ref{fig3} and the animations for these figures). Such different fates of F1 and F2 are probably due to the different angles between F1, F2, and F3.

Figure \ref{fig8} shows the temporal evolutions of the LOS magnetic field in the two regions where filament F3 rooted, before the flare. The left and right foot areas correspond to the network region and quiet Sun, respectively (Fig. \ref{fig8} a1--a4 and b1--b4). During a period of about 2 hours before the flare, we see there is no obvious flux emergence or cancellation in both areas. There is no approaching or separation between the positive and negative magnetic fields in the left foot area (Fig. \ref{fig8}c and d). The right foot area is mainly associated with the positive magnetic field; we see no obvious converging or diffusing motions of the field. In summary, there are no significant photospheric changes in the two foot areas of F3 before the flare. That means the large-scale and high-lying filament F3 is unlikely to erupt because of photospheric changes in this period of time. Therefore, the pushing of F1 and F2 is an important process for the eruption of F3. 

\begin{sloppypar}
Figure \ref{fig9} shows CME resulting from the multi-filament eruptions. Interestingly, the eruptions occur at the northwest of the solar disk close to the limb, yet they produce a halo CME with a velocity that reaches 1170.0 km/s (\url{https://cdaw.gsfc.nasa.gov/CME\_list/UNIVERSAL\_ver1/2023\_02/univ2023\_02.html}).  The CME results from the eruption of the large-scale  filament F3 and the combined pushing by filaments F1 and F2, which makes F3 to obtain a significant amount of initial kinetic energy in a very short period of time. This CME also drives a shock wave, which can effectively accelerate electrons and other particles. It is noteworthy that this halo CME triggered a geomagnetic storm (\url{https://www.swpc.noaa.gov/news/strong-g3-geomagnetic-storms-possible-february-27}).
\end{sloppypar}

Figure \ref{fig10} is a schematic diagram that illustrates the basic idea of a large-scale filament, crossing over an active region, is forced to erupt by a small-scale active-region filament pushing it out.  In this picture, two  types of filaments are involved: a large-scale intermediate or quiet filament carrying a large amount of material, which is usually stable and difficult to erupt, and an active-region filament, which is usually small but very active. The directions of the two filaments are almost coincident on the horizontal plane and almost parallel on the vertical plane. Their magnetic field directions are also similar. When the small-scale active-region filament erupts and begins to rise, it will then push the large-scale filament above to erupt. In this process, the collision, squeezing, and reconnection of the magnetic field lines between the surface layers of two filaments may occur, as in our observations in Figures \ref{fig3} and \ref{fig4}. This figure suggests that we should pay special attention to large-scale filaments suspended above an active region, as they could be triggered into eruption by underlying activity and eject a significant amount of material into the interplanetary space, posing significant hazards to our near-Earth environment.

\section{Summary and Discussion} 
      \label{S-Summary}      
      
We present observations on a multi-filament eruption which is associated with an M6.4 flare and a halo CME. This event gives us a unique insight into the dynamics of multi-filament eruptions and their corresponding effects on the interplanetary space.

Three filaments are involved in this eruption: two active-region filaments F1, F2, and a large-scale filament F3. They present a unique configuration: the two active-region filaments are located along the PIL, with F2 riding over F1, which coincides with the ``double-decker" model \citep{Liu2012}; the higher-altitude large-scale filament F3 extends crossing over the active region, lying along the direction of the PIL (Fig. \ref{fig2} and \ref{fig7}). Considering GOES X-ray flux and multi-wavelength observations, this event can be divided into three stages: 1) the eruptions of F1 and F2; 2) the interactions between the erupting F1 and F2 with F3; 3) the eruption of F3 (Fig. \ref{fig6}). 

We observe that the eruption of F3 is a completely forced process, being pushed out by the erupting F1 and F2.  This process has not been described and included in the conventional mechanisms of filament eruptions,  namely the MHD instabilities \citep[e.g.][]{Hood1981, Ji2003, Fan2005, Torok2005, Demoulin2010} and magnetic-reconnection-related processes \citep[e.g.][]{Antiochos1999, Chen2000, Lin2000, Moore2001}. In the traditional picture,  the filament eruption is caused by either the instability or triggered by magnetic reconnection occurring between the filament and surrounding magnetic field. The initial kinetic energy of the filament eruption comes from their own magnetic system. However, in our case study, the initial kinetic energy of F3 comes from the erupting F1 and F2 and the pushing process of F1 and F2 on F3, which lasts for more than 15 minutes.

There are also models on the interaction of two (or multi) flux ropes, such as the tilt-kink instability  \citep[e.g.][]{Keppens2014} and coalescence instability \citep[e.g.][]{Tajima1985, Murtas2021}. When two adjacent parallel flux loops develop opposite currents, the tilt-kink instability will occur. This instability has been suggested to be involved in the eruption that triggers flux ropes with a ``double-decker" configuration \citep[]{Keppens2014}. The coalescence instability could result in impulsive energy release with fast magnetic reconnection associated with the merging of two current carrying flux loops \citep{Tajima1985, Murtas2021}. In this study, we cannot conclude on the currents carried by the filaments and we do not observe clear signatures of strong reconenction between fialments F1 and F2 with filament F3. So, we are not sure whether the tilt-kink or coalescence instability would occur during the interactions between filaments F1 and F2 with filament F3. Nevertheless, any of them could still be possible.

\citet{Zhu2015} reported a ``double-decker" filament eruption with a contact angle of about $20^\circ$ between the two filaments. Their observations showed that the two filaments merged as the lower one erupted and collided with the upper one. In this process, a large amount of energy was released due to the reconnection between the magnetic fields of the two filaments. Their ``double-decker" filament eruption ultimately resulted in an M2.9 flare and a CME. Here, our study presents a different story. No significant or violent magnetic reconnection is observed, between the lower two active-region filaments F1, F2, and the upper large-scale filament F3, possibly due to their unique magnetic configurations. F1 and F3 are almost parallel to each other and are nearly on the same vertical plane with the same magnetic field direction. Therefore, strong magnetic reconnection is unlikely to occur between them, although there is expansion and untwisting motions during the rise of F1 eruption.

When F2 is pushing F3, a bright structure moving upwards along the left shoulder of F3 is evident in the high-temperature AIA wavelengths (Fig. \ref{fig4}). This bright structure should come from the heating of the plasma by squeezing and magnetic reconnection between the top of F2 and F3. Since the uplift process of F2 is also accompanied by untwisting, it is difficult to determine how large the contact angle is between the magnetic fields of the top of F2 and the left shoulder of F3. We cannot judge the specific contribution of magnetic reconnection to the plasma heating. However, even if partial magnetic reconnection occurs between them, it is not enough to destroy the overall magnetic structure of F3. As we see, F3 is eventually pushed out to eruption nearly without being restructured. This is evidently different from the observations of \citet{Zhu2015}. 

An interesting question is whether F3 could erupt spontaneously without the pushing from F1 and F2. F3 is a large-scale filament with a length of about 360 Mm. We see a lot of cold and dense plasma gathered in the central dip of the filament (Fig. \ref{fig1}, \ref{fig2}, and \ref{fig7}). It may take a long time to heat and evaporate the plasma in the filament to reduce gravity.  It is also difficult for the plasma within the filament to be spontaneously transported down to the lower atmosphere through the two legs of the filament by some kind of disturbance. As we see, even though the eruptions of F1 and F2 have already started for several minutes, F3 still remained stable, while the pushing by F1 and F2 on F3 persisted for more than 15 minutes to make the F3 finally erupt.

It is generally believed that the magnetic topology of a filament plays a key role in determining whether it will erupt or not \citep{Chen2020}.  However, in our case it is difficult to derive the magnetic structure of F3 through extrapolation. F3 is very large and rooted in a region of weak magnetic field. Being close to the limb also brings in a significant projection effect.  For the photospheric magnetic field in the regions where F3 rooted, we do not see any shearing, rotating, converging, or diffusing motions (Fig. \ref{fig8}). This means that the magnetic topology of F3 should remain unchanged without external disturbances. This makes F3 to be a relatively stable filament. We speculate that F3 would not erupt, at least in a short time period, without the pushing from F1 and F2.

Our study provides an important indication on the impact of combined eruptions on space weather. Active-region filaments are very active and easy to erupt. However, their lengths are limited to the size of the active region, thus further restricting the scale of their eruptions. Quiet and intermediate filaments are large in size and, so, their mass is significant, but they are stable and difficult to erupt. If a large-scale filament is just suspended above an active region with a small-scale filament below it, as in our study, when the lower active-region filament erupts, it will then push the upper one to erupt (Fig. \ref{fig7} and \ref{fig10}). This means that a large amount of plasma will be ejected into the interplanetary space at a high initial velocity, resulting in a large space weather influence. So our study suggests that we should pay particular attention to configurations where a large-scale filament suspends above an active region with one or more underneath active-region filaments. This type of topology has a great potential to result in severe space weather events.


\begin{acks}
 We are grateful to Dr. Hechao Chen at Yunnan University, Dr. Yijun Hou at National Astronomical Observatories of Chinese Academy of Sciences (CAS) and Dr. Zhenyong Hou at Peking University for their valuable discussions. The authors thank the ASO-S, SUTRI, CHASE, SDO, STERO and GOES teams for providing the data. ASO-S mission is supported by the Strategic Priority Research Program on Space Science of CAS (Grant No. XDA15320000). The CHASE mission is supported by China National Space Administration (CNSA). SDO is a space mission in the Living With a Star Program of NASA. SOHO is a project of international cooperation between ESA and NASA. SUTRI is a collaborative project conducted by the National Astronomical Observatories of CAS, Peking University, Tongji University, Xi’an Institute of Optics and Precision Mechanics of CAS and the Innovation Academy for Microsatellites of CAS. The CME catalog used is generated and maintained at the CDAW Data Center by NASA and The Catholic University of America in cooperation with the Naval Research Laboratory. 
 
\end{acks}

\begin{authorcontribution}
Y.L. Song and J.T. Su carried out this study. 
Q.M. Zhang analyzed the associated CME. 
M. Zhang and Y.Y. Deng provided suggestions on the design of study. 
X.Y. Bai, S. Liu, X. Yang, J. Chen, H.Q. Xu, K.F. Ji and Z.Y. Hu provided suggestions on the data analysis.
Y.L. Song wrote the manuscript. 
J.T. Su and M. Zhang revised the manuscript.
All the authors helped to improve the manuscript.
\end{authorcontribution}

\begin{sloppypar}
\begin{fundinginformation}
This work is supported by National Key R\&D Program of China (Grant No. 2022YFF0503001, 2021YFA0718601), National Natural Science Foundation of China (Grant No. 12173049, 11973056, 11803002,) and Beijing Natural Science Foundation (Grant No. 1222029).
\end{fundinginformation}
\end{sloppypar}

 \begin{dataavailability}
The ASO-S data used in this study are not publicly available, since they were observed during the commissioning phase (from 9th October, 2022 to 31st March, 2023). But for reasonable request, they can be available from the corresponding authors. SDO, SUTRI data used in this work are publicly accessible on the websites (\url{http://jsoc.stanford.edu/ajax/lookdata.html} and \url{https://sun10.bao.ac.cn/SUTRI/}). The data of STEREO-A and SOHO can be obtained through the Virtual Solar Observatory (\url{https://sdac.virtualsolar.org/cgi/search}). The H$\alpha$ data from CHASE and GONG can be accessed from their official websites (\url{https://ssdc.nju.edu.cn/home} and \url{https://gong2.nso.edu/archive/patch.pl?menutype=s}).
 \end{dataavailability}



\begin{conflict}
The authors declare that they have no conflicts of interest.
\end{conflict}


\bibliographystyle{spr-mp-sola}
\bibliography{sola_bibliography}

\end{document}